\documentclass[10pt,preprint]{aastex}

\slugcomment{Accepted for publication in ApJL}

\shorttitle{Travel-Time Signature of Chromospheric Flows}
\shortauthors{Nagashima et al.}

\usepackage{natbib,graphicx}
\usepackage{color, graphicx}

\begin{document}


\title{Helioseismic Signature of Chromospheric Downflows in Acoustic 
Travel-Time Measurements from \protect\textit{Hinode}}


\author{Kaori Nagashima\altaffilmark{1}, Takashi Sekii\altaffilmark{1,2},
Alexander G. Kosovichev\altaffilmark{3}, Junwei Zhao\altaffilmark{3}, and
Theodore D. Tarbell\altaffilmark{4}}

\altaffiltext{1}{Department of Astronomical Science, 
The Graduate University for Advanced Studies (Sokendai), \\
National Astronomical Observatory of Japan, 2-21-1 Osawa, 
Mitaka, Tokyo 181-8588, Japan.}
\email{kaorin@solar.mtk.nao.ac.jp}
\altaffiltext{2}{National Astronomical Observatory of Japan, 
2-21-1 Osawa, Mitaka, Tokyo 181-8588, Japan.}
\altaffiltext{3}{W.W. Hansen Experimental Physics Laboratory, 
Stanford University, Stanford, CA 94305, USA.}
\altaffiltext{4}{Lockheed Martin Solar and Astrophysics Laboratory, 
B/252, 3251 Hanover St., Palo Alto, CA 94304, USA.}



\begin{abstract}
We report on a signature of chromospheric downflows 
in two emerging-flux regions detected by time--distance helioseismology analysis.
We use both chromospheric intensity oscillation data in the Ca \textsc{ii} H line
and photospheric Dopplergrams in the Fe \textsc{i} 557.6nm line  
obtained by \textit{Hinode}/SOT for our analyses.
By cross-correlating the Ca \textsc{ii} oscillation signals, we have
detected a travel-time anomaly in the plage regions;
outward travel times are shorter than inward travel times by $0.5 - 1$ minute.
However, such an anomaly is absent in the Fe \textsc{i} data.
These results can be interpreted as evidence of downflows in the lower chromosphere.
The downflow speed is estimated to be below $10 \ \mathrm{ km \  s^{-1}}$.
This result demonstrates a new possibility of studying chromospheric flows
by time--distance analysis.
\end{abstract}





\section{INTRODUCTION}

Local helioseismological techniques have been exploited to reveal
subsurface structures and dynamics of the Sun,
for instance, mass flows and wave-speed variations below sunspots 
(e.g., \citealt{2001ApJ...557..384Z}), or
subsurface structure of supergranular flows (e.g., \citealt{2007PASJ...59..637S}).
In this Letter, we examined whether we can 
detect flows in emerging magnetic flux regions.

Magnetic flux tubes are generated in the convection zone.
They emerge into the photosphere due to magnetic buoyancy,
forming sunspots and/or restructuring the coronal magnetic configuration,
eventually leading to activity phenomena, such as flares.
Emerging flux regions are the place where such flux tubes 
are first observed directly;
their configurations and dynamics are manifestations
of interaction between plasma and magnetic flux
in the subphotosphere. Thus, studying various aspects of flux emergence is 
vitally important in understanding solar and stellar dynamo mechanisms.

Numerous studies about emerging flux regions have been done both 
observationally and theoretically, and revealed complicated structures and
flow patterns
(e.g., \citealt{1996A&A...306..947S}; \citealt{2000SoPh..192..159K};
\citealt{2006PASJ...58..407K}; \citealt{2006ASPC..354..249P};
 \citealt{2008ApJ...685L..91M}; \citealt{2008ApJ...687.1373C}).
It is generally thought that along the flux tube 
are plasma downflows.
The evacuated plasma decelerates as it descends due to the increasing density;
in fact, the observed speed is $30-50 \ \mathrm{km\ s^{-1}}$ in the chromosphere and
$1-2 \ \mathrm{km\ s^{-1}}$ in the photosphere 
(e.g., \citealt{2002plap.book.....T} and references therein).
To understand how emerging flux regions evolve and 
how they affect activity phenomena,
it is essential to study them in a range of layers 
from the subphotosphere to the upper atmosphere.

Local helioseismology is the only way to investigate the subphotospheric 
structure and dynamics of the Sun.
Moreover, it allows us to study the dynamics above the visible surface.
In this Letter, we report on a signature of chromospheric flows detected 
by a time--distance helioseismology analysis.
Examining two emerging flux regions observed by \textit{Hinode}, 
we detect a signal in the travel-time differences, which may be caused by
plasma downflows in the chromosphere.
Another implication of the present work is that 
chromospheric dynamics must be taken into account when 
chromospheric lines are used in helioseismology analysis.
We briefly introduce \textit{Hinode} observations in Section \ref{sec:obs},
and describe the analysis methods and results in Section \ref{sec:td}.
Discussion of our interpretation and conclusion are given in Section \ref{sec:dis}.

\section{OBSERVATIONS} \label{sec:obs}

The Solar Optical Telescope (SOT; \citealt{2008SoPh..249..167T})
onboard the \textit{Hinode} satellite \citep{2007SoPh..243....3K}
made two observing runs for helioseismology
around the solar disc centre:
on 2007 November 23 (11:17--23:55 UT) and
on 2007 December 04 (0:14--23:58 UT).
Hereafter, we refer to the first 12-hour run as data set 1,
and the second 24-hour run as data set 2.
In the field of view of data set 1, 
an emerging flux region appeared before the observation started,
showing a plage and small sunspots. 
This region was registered as NOAA Active Region 10975.
It disappeared within two days after the \textit{Hinode} observations.
In the field of view of data set 2, a decaying active region 
(NOAA Active Region 10976) was observed.
Sample images of these regions are shown in Figure \ref{fig:samples}.

During these observations,
both SOT instruments, the Narrowband Filter Imager (NFI) and the Broadband Filter Imager (BFI),
were used: the BFI took Ca \textsc{ii} H images, while the NFI took
nonmagnetic Fe \textsc{i} 557.6 nm filtergrams.
The time cadence was 60 seconds.
The Fe \textsc{i} Dopplergrams are calculated from 
the blue-wing and the red-wing intensities of the Fe \textsc{i} 557.6 nm line;
that is, the (blue -- red) signal is divided by the (blue + red) signal.
During the each observing period \textit{Hinode} tracked the regions 
using the correlation tracker (CT; \citealt{2008SoPh..249..221S}) 
to stabilize the images.
The field of view is 218 $\times$ 109 arcsec for the Ca \textsc{ii} H data 
and 328 $\times$ 164 arcsec for the Fe \textsc{i} data,
while the pixel sizes are 0.108 arcsec (for Ca \textsc{ii} H) 
and 0.16 arcsec (for Fe \textsc{i}).
In this analysis, we use only the field of view of the Ca \textsc{ii} H data.
The Fe \textsc{i} data are interpolated to the Ca \textsc{ii} H data grid.
We also apply $2\times 2$ pixel summing, increasing the pixel scale to about 0.2 arcsec.
For the intensity data sets, we use running difference of images to 
remove possible spatial trend.

\section{TIME--DISTANCE ANALYSIS AND ITS RESULTS}  \label{sec:td}

We analyze the 12-hour and 24-hour data sets (data sets 1 and 2) by a 
time--distance helioseismology technique \citep{1993Natur.362..430D}.
First, we apply three filters in the frequency--wavenumber space:
(1) an f-mode filter to remove the f-mode signal, since we use p-mode waves only,
(2) low-wavenumber filters and a Gaussian frequency filter 
with a $1$-mHz width peaking at $3.3 \ \mathrm{ mHz}$
to remove the known artifacts (see \citealt{2008SoPh..249..167T})
and select the main p-mode signal,
(3) a phase-speed filter with a width of $4.4  \ \mathrm{ km \ s^{-1}}$
centered at a speed of $22 \ \mathrm{ km \ s^{-1}}$ to select 
waves travelling to a distance of $14.86 \ \mathrm{ Mm}$,
which is chosen for this study.

Second, by cross-correlating the filtered signals,
we measure the travel-time difference between the outward and 
inward wave components following the standard time--distance helioseismology procedure
\citep{1997ASSL..225..241K}.
For this, we average the signals over an annulus 
of the radius of 14.86 Mm around a target
point, and cross-correlate the signals at the target point and that of the annulus.
We measure outward (from the target to annulus) and 
inward (from annulus to the target) travel times
at each point in the field of view by fitting 
a Gabor-wavelet function to the
cross-correlation function.
The fitting function, $C(\Delta, \tau)$, 
for spatial displacement $\Delta$, and time lag $\tau$ has the form
\begin{eqnarray}
C(\Delta, \tau) = A \cos(\omega_0 (\tau-\tau_\mathrm{ p}) ) 
\exp{\left[-\left(\frac{\delta \omega}{2} (\tau - \tau_\mathrm{ g})\right)^2\right]} ,
\end{eqnarray}
where 
$A$ is the amplitude, $\omega_0$ is the central frequency,
$\delta \omega$ is the frequency bandwidth, and
$\tau_\mathrm{ g}$ and $\tau_\mathrm{ p}$ are the group and phase travel times.
These five parameters are determined by applying a nonlinear least-squares fitting 
method (Levenberg--Marquardt method; see, e.g., \citealt{1992nrfa.book.....P}). 
Errors of the fitted parameters are derived from the estimated covariance matrix.
It should be mentioned that
we used the phase travel time in the following discussions, because
the phase travel time is determined with higher precision
than the group travel time.

Figure \ref{fig:tmaps} shows outward--inward travel-time difference maps.
In these maps, misfitted points are assigned null travel-time differences;
the number of misfittings is insignificant.
In the quiet regions, both the Ca \textsc{ii} H intensity data and 
Fe \textsc{i} Doppler data show the same pattern of 
flow divergence due to supergranulation.
This is similar to the results from a previous \textit{Hinode} observation
reported by \citet{2007PASJ...59..637S}.
In the plage region, we find a strong travel-time anomaly 
in the Ca \textsc{ii} H data:
the outward travel time is shorter than the inward travel time by about 
0.5 - 1 \ minute on average.
This anomaly is not present in the Fe \textsc{i} $557.6 \ \mathrm{ nm}$ Doppler data.
We also analyze the Fe \textsc{i} $557.6 \ \mathrm{ nm}$ intensity data
from the same observing runs.
These data are noisier, but they show the travel-time difference maps 
consistent with these obtained from the Fe \textsc{i} $557.6 \ \mathrm{ nm}$ 
Dopplergrams. Particularly, they do not show the travel-time anomaly, either.

The anomaly in the outward--inward travel-time difference in the chromospheric data 
is seen also in the averaged cross-correlation function.
We average the cross-correlation functions
in the quiet Sun (QS) region and in the plage for both data sets 1 and 2.
We define the plage as a region where the Ca \textsc{ii} H intensity is $25 \%$
larger than the average intensity of QS.
By fitting the Gabor wavelet to the averaged cross-correlation functions,
we obtain the travel times listed in Table \ref{tbl:avtime}.
We find that in the plage the outward travel time is 
about $0.67 \ \mathrm{min}$ (for data set 1) or $0.42 \ \mathrm{min}$ (for data set 2) 
shorter than the inward travel time only in the chromospheric Ca \textsc{ii} H data,
while in the QS 
the difference between the outward and inward travel times is insignificant
in both the chromospheric and the photospheric data.

We also have checked temporal variation of the travel-time anomaly
for data set 2. The 24-hr run is divided into three 8-hr runs,
and the travel times are measured for each 8-hr run.
As a result, we obtain the following sequences of the outward--inward travel-time 
difference in the chromospheric Ca \textsc{ii} H data:
$(-5.35 \pm 0.04) \times 10^{-1} \ \mathrm{min}$, 
$(-4.16 \pm 0.04) \times 10^{-1} \ \mathrm{min}$, and
$(-4.21 \pm 0.06) \times 10^{-1} \ \mathrm{min}$  in the plage region, and
$(-1.11 \pm 0.03) \times 10^{-1} \ \mathrm{min}$, 
$(-0.14 \pm 0.07) \times 10^{-1} \ \mathrm{min}$, and
$(+1.37 \pm 0.03) \times 10^{-1} \ \mathrm{min}$ in the QS.
In the plage, for all the 8-hr runs the anomaly similar to the one 
detected in the entire run is detected again, and 
particularly it is rather stronger in the first 8-hr run.
This variation is consistent with a decaying behaviour of the plage region
if we interpret that the travel-time anomaly results from a chromospheric downflow
(see Section \ref{sec:dis} for details), and that
the downflow weakened during the observation.
We also find, in the developing plage in data set 1, the anomaly 
being stronger in the latter half of the observation.
In the QS, the weak travel-time anomaly simply corresponds to the 
supergranular pattern. The observed variation with time implies that
the lifetime of supergranules is slightly less than 1 day.
This is consistent with previous studies
(e.g., \citealt{2002tsai.book.....S}).

\section{DISCUSSION AND CONCLUSION}  \label{sec:dis}

In our analysis, we have used only acoustic (p-mode) waves with the central frequency of 
$3.3 \ \mathrm{mHz}$.
Although stationary harmonic p-mode waves at this frequency are evanescent in the chromosphere,
in reality such perturbations excited by impulsive subphotosphere sources 
do propagate upward (e.g., \citealt{1975hydr.book.....L}; \citealt{1992ApJ...387..707G}).
This certainly requires further careful modelling, but here for the initial interpretation we assume that 
the waves travel roughly with the chromospheric sound speed.
Thus, a chromospheric signal is observed after the corresponding photospheric signal 
is observed and the time delay is the acoustic travel time between the two layers,
which is estimated by $L/c \sim 25 \ \mathrm{second}$, where 
$L \sim 250$\,km is the formation height difference between the 
Fe \textsc{i} 557.6nm line and the Ca \textsc{ii} H line 
\citep{2007PASJ...59S.663C} and
$c \sim 10$\,km\,s$^{-1}$ is the sound speed in the chromosphere.
Indeed, the phase shift between the photospheric G-band intensity signal
and chromospheric Ca \textsc{ii} intensity signal around 4 mHz on the p-mode ridge,
reported by \citet{2008A&A...481L...1M}, implies a $\sim 30$-second propagation time
between the photosphere and the chromosphere.
Such a direct measurement in our case is not straightforward
because our photospheric signal is Doppler velocity.
Understanding phase shifts between Doppler signal and intensity signal would
be part of a future work.
Please note, however, the travel-time analysis we use is not affected by such a difficulty
as we cross-correlate the same observables.

\begin{figure}
\includegraphics[bb=0 0 144 145, width=0.25\textwidth]{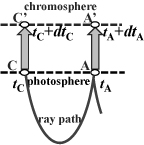}
\caption{Schematic illustration showing
wave propagation (ray path) from the subphotosphere to the chromosphere.
If the wave arrival times at A and C in the photosphere are $t_\mathrm{A}$ and
$t_\mathrm{C}$, the disturbance reaches at A$^{\prime}$ and C$^{\prime}$ in the chromosphere 
after some time delay, i.e.,
at $t_\mathrm{A} + dt_\mathrm{A}$ and $t_\mathrm{C}+dt_\mathrm{C}$
(see the text in Section \ref{sec:dis}).
\label{fig:path}}
\end{figure}

Figure \ref{fig:path} shows how the disturbance caused by a source at point C
propagates. If the wave arrival times at points A and C in the photosphere
are $t_\mathrm{ A}$ and $t_\mathrm{ C}$,
the travel time from C to A is simply 
 $t_\mathrm{C\, to\, A} =t_\mathrm{ A}-t_\mathrm{ C}$.
The `travel time' from C$^{\prime}$ to A$^{\prime}$,
defined through the cross-covariance fitting procedure,
$t_\mathrm{C^{\prime} \, to\,  A^{\prime}}$, is identical\, to\, $t_\mathrm{C\, to\, A}$
if the time delays at A, $dt_\mathrm{ A}$, and at C, $dt_\mathrm{ C}$,
are the same.
Our travel-time measurements in the QS are consistent with this picture;
the photospheric and the chromospheric travel times are 
essentially the same (see Table \ref{tbl:avtime}).

If there is a downflow above C, however, and there is no downflow above A,
then the time delays at two points are different: $dt_\mathrm{ C} > dt_\mathrm{ A}$.
By setting point C as the center of an annulus, or as the target point,
and point A as a point on the annulus around the target point,
we can describe the outward travel time at the photospheric level, $t_\mathrm{ out, ph}$, and
the travel time at the chromospheric level, $t_\mathrm{ out, ch}$, as
\begin{eqnarray}
 t_\mathrm{out, ph}&=&t_\mathrm{ C\, to\, A} =t_\mathrm{A} - t_\mathrm{C} \\ 
 t_\mathrm{out, ch} &=&t_\mathrm{ C^{\prime}\, to\, A^{\prime}} =t_\mathrm{ C\, to\, A} +(dt_\mathrm{ A} -dt_\mathrm{ C}) < t_\mathrm{ C\, to\, A} \  ,  \label{eq:t_out_ch}
\end{eqnarray}
and also the corresponding inward travel times $t_\mathrm{ in, ph}$ and  $t_\mathrm{ in, ch}$ as
\begin{eqnarray}
 t_\mathrm{in, ph}&=& t_\mathrm{ A\, to\, C} = t_\mathrm{ C\, to\, A} \\
 t_\mathrm{in, ch}&=&t_\mathrm{ A^{\prime}\, to\, C^{\prime}} = t_\mathrm{ A\, to\, C}-(dt_\mathrm{ A} -dt_\mathrm{ C})> t_\mathrm{ A\, to\, C} \ , \label{eq:t_in_ch}
\end{eqnarray}
where we assume for simplicity that there is no perturbation in the subphotospheric layer,
or, equivalently, removed the subphotospheric components from the discussion.
Then, the outward--inward travel-time difference in the chromosphere,
$\Delta t_\mathrm{ch}$, is
\begin{eqnarray}
\Delta t_\mathrm{ch}=t_\mathrm{out, ch}-t_\mathrm{in, ch} =2 (dt_\mathrm{ A} - dt_\mathrm{ C}) <0 . \label{eq:dis_tout_tin}
\end{eqnarray}
The time delay difference between the two points $dt_\mathrm{ A} - dt_\mathrm{ C}$ 
can be estimated by
\begin{eqnarray}
dt_\mathrm{ A} - dt_\mathrm{ C} = -\frac{L}{c} \frac{V}{c}, \label{eq:dis_dt}
\end{eqnarray}
where $V$ is the downflow speed.
Using Equations (\ref{eq:dis_tout_tin}) and (\ref{eq:dis_dt}),  
we can roughly estimate the downflow speed:
at $V \sim 8 \  \mathrm{ km \ s^{-1}}$ for the plage in data set 1 
($\Delta t_\mathrm{ch}\sim  -0.67 \ \mathrm{ minute }$) and 
at $V \sim 5 \ \mathrm{ km \ s^{-1}}$ for the plage in data set 2
($\Delta t_\mathrm{ch}\sim  -0.42\  \mathrm{ minute }$).

Note that Equatios (\ref{eq:t_out_ch}) and (\ref{eq:t_in_ch}) imply that the
chromospheric mean travel time $(t_\mathrm{out,ch} + t_\mathrm{in,ch})/2$
should essentially be identical to the photospheric mean travel time
$(t_\mathrm{out,ph} + t_\mathrm{in,ph})/2$.
Table \ref{tbl:avtime} shows that this is indeed the case.

\begin{figure}
\includegraphics[bb=0 0 414 188, width=0.7\textwidth]{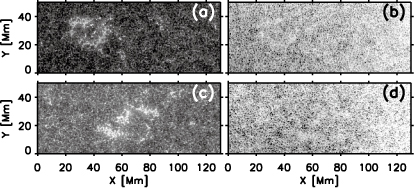}
\caption{Sample images from two data sets.
{\it Top row:} plage region observed on 2007 November 23 (data set 1),
Ca \textsc{ii} H line intensity image (a) and 
Fe \textsc{i} $557.6 \ \mathrm{ nm}$ Dopplergram (b).
{\it Bottom row:} eecaying region observed on 2007 December 04 (data set 2),
Ca \textsc{ii} H intensity image (c) and Fe \textsc{i} Dopplergram (d).
In the Dopplergrams, white means a redshifted signal, 
while black means a blueshifted signal.
\label{fig:samples}}
\end{figure}

\begin{figure}
\includegraphics[bb=0 0 414 188, width=0.7\textwidth]{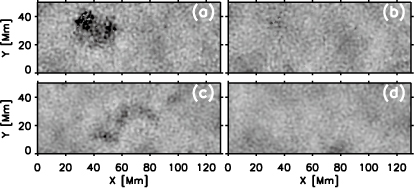}
\caption{Outward--inward travel-time difference maps.
{\it Top row:} data set 1,
(a) Ca \textsc{ii} H line intensity and (b) Fe \textsc{i} $557.6 \ \mathrm{nm}$ (Dopplergram).
{\it Bottom row:} data set 2,
(c) Ca \textsc{ii} H line intensity and (d) Fe \textsc{i} $557.6 \ \mathrm{nm}$ (Dopplergram).
The gray scale corresponds to the range from $-1 \ \mathrm{ minute}$ to  $+1 \ \mathrm{ minute}$.
The field of view is the same as in Figure \ref{fig:samples}. \label{fig:tmaps}}
\end{figure}

Actually, we do detect the photospheric downflows in the plage region
in the photospheric Fe \textsc{i} Dopplergram. 
Presumably these flows are much weaker than the flows in the chromosphere.
Although we still need a more 
fine-tuned calibration of the NFI data to derive the better estimates of the speed,
a redshift signal is clearly seen in the Dopplergram (see Figure \ref{fig:samples}).
Since the signal is somewhat comparable though weaker than the Doppler shift due to the solar rotation,
the downflow speed is roughly estimated at a few hundred $\mathrm{m \ s^{-1}}$.
This is consistent with 
the deceleration picture in emerging flux regions.

\begin{figure}
\includegraphics[bb=0 0 331 159, width=0.35\textwidth]{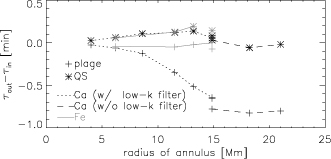}
\caption{
Outward--inward travel-time difference for data set 1 plotted against 
annulus size. The crosses and asterisks indicate the difference measured in plage and QS,
respectively. The gray solid lines indicate the differences for Fe \textsc{i} Dopplergram.
The dotted and dashed lines are for the Ca \textsc{ii} intensity data; 
the former and the latter are the differences measured with and without low-wavenumber 
filters (see the text for the details).
Original values are also shown at 14.86 Mm, which are measured using a
frequency filter different from others, although the values with
different filters are essentially the same.
Error bars are smaller than the symbol size.
\label{fig:rt}}
\end{figure}

In our analysis, we use $14.86$-Mm annulus to measure the inward and outward 
travel times. We chose this size because we know that this size of annulus
is suitable for detecting spatial patterns of the size of
the supergranular flow, as is mentioned 
in the previous works by \citet{2007PASJ...59..637S}.
We also check the cases for different sizes of annulus for data set 1: from 4 Mm to around 20 Mm.
Phase-speed filters are changed depending on the annulus size.
For these calculations,
so that we can measure travel times for smaller annuli,
we use a Gaussian frequency filter 2.4-times broader than that
mentioned in Section 3, although it renders wavepackets narrower, and  makes 
the phase travel-time measurement less precise.
For the case of annuli larger than the original one, 
the low-wavenumber filter, required for removing the artifacts in Fe \textsc{i}
data, cannot be used. Thus, for these annuli the travel-time can be measured only 
for the Ca \textsc{ii} data.
Results are shown in Figure \ref{fig:rt}.
In the QS, the travel-time difference remains small regardless of the size, 
while the difference in plage increases as the annulus size increases.
This can be explained 
by increasingly large part of an annulus located outside plage 
where there is no downflow, which contributes to the travel-time anomaly.
In fact, when the size attains the spatial extent of the plage ($\sim 15-20 \ \mathrm{Mm}$),
the anomaly seems to reach a plateau.

We note that a sound speed anomaly in the chromosphere can also
lead to an outward--inward travel-time difference through affecting
the chromospheric time delay.
Such an anomaly may arise from magnetic field. However, generally in plage 
magnetic field is not strong enough to explain the travel-time difference 
we have found. Indeed, in MDI \citep{1995SoPh..162..129S} magnetogram 
the plage magnetic field in data set 1 is a few hundred Gauss,
which translates to only a few hundred $\mathrm{m\ s^{-1}}$ 
(magneto-)sound-speed anomaly, and in data set 2 it is even weaker.
A temperature increase in the plage does not explain the travel-time
difference either, since it will \textit{decrease} the time delay rather
than increase it.

Let us also consider the effect of horizontal motions in the plage.
The plage regions changed their shape and the bright patches and/or the 
small sunspots in the field of view were moving during the observation periods.
The speed of these motions was $\sim 1 \ \mathrm{km \ s^{-1}}$ at most, and
this is slower than the sound speed by an order of magnitude.
Our estimates show that these motions are too slow to affect the estimates 
of the chromospheric downflows in any way.

In this study, we have found a signature of chromospheric downflows 
in the differences between the acoustic travel times measured simultaneously in the 
photosphere and chromosphere.
The outward--inward travel-time anomaly is found only 
in the chromospheric measurements of a plage region, and, therefore,
this can be interpreted as chromospheric downflows in the plage. 
This kind of signature reminds us that
chromospheric helioseismology data may include not only information about subsurface structures 
but also about chromospheric structures and dynamics.
This needs to be taken into account in analysis and interpretation of
helioseismic observations in the chromosphere.
Such effects have not been sufficiently studied.

On the other hand, this result opens new possibilities of 
multiwavelength time--distance helioseismology studies of 
chromospheric flows and structures.
As recent high-resolution observations have revealed, 
the chromosphere is full of activities and flows (e.g., \citealt{2007Sci...318.1594K}).
Multiwavelength helioseismic observations may provide a
 new method for detecting chromospheric flows, which may be of great 
importance for revealing dynamics of the chromosphere.



\acknowledgments

We thank Tetsuya Magara for helpful comments.
\textit{Hinode} is a Japanese mission developed and launched by ISAS/JAXA, 
with NAOJ as domestic partner and NASA and STFC (UK) as international partners. 
It is operated by these agencies in co-operation with ESA and NSC (Norway). 
This work was carried out at the NAOJ Hinode Science Center,
which is supported by the Grant-in-Aid for 
Creative Scientific Research 
``The Basic Study of Space Weather Prediction'' from MEXT, Japan 
(Head Investigator: K. Shibata), generous donations 
from Sun Microsystems, and NAOJ internal funding.
K.~Nagashima is supported by the Research Fellowship from the 
Japan Society for the Promotion of Science for Young Scientists.



{\it Facilities:} \facility{{\it Hinode} (SOT)}

\begin{deluxetable}{rccccc}
\tablecolumns{6}
\tablewidth{0pc} 
\tablecaption{Travel Times (in minutes) 
for Various Data Sets, Regions and Spectral Lines \label{tbl:avtime}}
\tablehead{
\colhead{} & \multicolumn{2}{c}{Data Set 1} & \colhead{} &\multicolumn{2}{c}{Data Set 2} \\
\cline{2-3} \cline{5-6}\\
& Ca \textsc{ii} H &Fe \textsc{i} && Ca \textsc{ii} H &Fe \textsc{i} }
\startdata
\sidehead{Plage}
In &     22.043 $\pm$    2.8E-03 &     21.725 $\pm$    1.8E-03 & &    22.104 $\pm$    3.0E-03 &     21.788 $\pm$    1.8E-03 \\
 Out &     21.372 $\pm$    3.0E-03 &     21.646 $\pm$    2.0E-03 &&     21.685 $\pm$    3.4E-03 &     21.892 $\pm$    2.0E-03 \\
Out-In &  -6.72E-01 $\pm$    4.1E-03 &  -7.82E-02 $\pm$    2.7E-03 &&  -4.19E-01 $\pm$    4.6E-03 &   1.03E-01 $\pm$    2.7E-03 \\
\hline
\sidehead{QS}
         In &     21.964 $\pm$    2.1E-03 &     21.806 $\pm$    1.5E-03 &&     22.232 $\pm$    2.8E-03 &     22.019 $\pm$    2.2E-03 \\
      Out &     22.027 $\pm$    3.3E-03 &     21.939 $\pm$    2.1E-03 &&     22.249 $\pm$    3.2E-03 &     21.993 $\pm$    2.1E-03 \\
    Out-In &   6.34E-02 $\pm$    3.9E-03 &   1.33E-01 $\pm$    2.6E-03 & &  1.66E-02 $\pm$    4.3E-03 &  -2.55E-02 $\pm$    3.1E-03 \\
\enddata
\end{deluxetable}


\end{document}